\documentclass[11pt]{article}
\usepackage{moriond,epsfig,subfigure,floatflt}

\def\be{\begin{equation}}
\def\ee{\end{equation}}
\def\bea{\begin{eqnarray}}
\def\eea{\end{eqnarray}}

\begin{document}
\vspace*{4cm}
\title{HIGH $P_{T}$ SUPPRESSION AT FORWARD RAPIDITIES IN D+AU AND AU+AU AT $\sqrt{s_{NN}}=200$ GEV}

\author{C\u at\u alin Ristea for the BRAHMS collaboration}

\address{Niels Bohr Institute, University of Copenhagen, Denmark}

\maketitle\abstracts{
We present centrality dependent charged hadron yields at several 
pseudo-rapidities from Au+Au collisions at $\sqrt{s_{NN}}=200$GeV 
measured with BRAHMS spectrometers. Nuclear modification factors $R_{AA}$ and $R_{CP}$ 
for charged hadrons at forward angles in Au+Au and d+Au collisions at RHIC will be 
discussed.
}

\section{Introduction}

From the nucleon-nucleon interactions it is known that when 
two partons undergo a scattering with large momentum transfer Q$^2$ in the early 
stages of collision, the hard-scattered partons fragment into jets of hadrons
with high transverse momentum ($p_{T}>2$GeV/c)~\cite{pp}. When the hard scattered 
partons will traverse the hot and dense nuclear matter created in a high energy 
nucleus-nucleus collision, they lose energy through gluon bremsstrahlung with the energy 
lost depending on the density of color charges in the matter through they pass~\cite{gy,w2}.
This effect is called jet quenching and the most directly measurable
consequence is the suppression of high transverse momentum hadrons in the final state.
Therefore any modification in the high $p_{T}$ spectrum is probing the high density medium created
in the collision.

All four experiments at RHIC have been reported that the high $p_{T}$ inclusive hadron yields
in central Au+Au collisions are largely suppressed as compared to p+p or peripheral Au+Au
collisions, scaled by the number of contributing binary (nucleon-nucleon) 
collisions. In the midrapidity region such suppression was not seen in d+Au collisions at RHIC, 
indicating that it is a final state effect associated with the hot and dense matter 
produced in Au+Au collisions~\cite{phobos,phenix,star,brahms}.

The observed suppression cannot be explained by energy loss at the hadronic stage, but rather
by the existence of a high color field which leads to the depletion of high momentum particles.
The amount of energy loss is not so straightforward to deduce, because there could be
processes like modifications of the parton distribution functions and the scatterings of the incoming 
partons prior to the hard scattering, which are termed as initial state effects.
In order to disentangle the initial from final state effects, the BRAHMS experiment
study different collision systems p+p, d+Au, Au+Au, Cu+Cu, at different energies.


In this paper we are presenting preliminary results from Au+Au at 200 GeV from the
high statistics run in 2004, and also d+Au data, in the context of the above
processes.

In order to measure the high $p_{T}$ hadron suppression in nucleus-nucleus collisions, the
comparison of the hadron $p_{T}$ spectra relative to reference data from nucleon-nucleon
collisions at the same collision energy is needed. The nuclear modification factor is
defined as:

\begin{equation}
R_{AA}(p_{T})=\frac{d^{2}N/dp_{T}d\eta}{T_{AA}d^{2}{\sigma^{NN}}/dp_{T}d\eta}
\end{equation}
where $T_{AA}=<N_{bin}>/{\sigma^{NN}_{inel}}$ accounts for the collision geometry,
averaged over the event centrality class. $<N_{bin}>$, the equivalent number of binary
NN collisions, is calculated using the Glauber model. $\sigma_{inel}$ and
$d^{2}{\sigma^{NN}}/dp_{T}d\eta$ are the cross section and differential cross section
for inelastic nucleon-nucleon (NN) collisions, respectively. In the absence of nuclear
medium effects such as shadowing, the Cronin effect or gluon saturation, hard processes
are expected to scale with $<N_{bin}>$ and $R_{AA}$=1. Any deviation from unity
indicate nuclear medium effects.

In order to remove the systematic errors introduced by the comparison of the measurements
of nucleus-nucleus and p+p collisions, we construct the ratio of central to peripheral
collisions, $R_{CP}$, defined as:

\begin{equation}
R_{CP}=\frac{1/<N_{bin}^{C}>}{1/<N_{bin}^{P}>}\frac{dN^{C}/dp_{T}d{\eta}}{dN^{P}/dp_{T}d{\eta}}
\end{equation}
where $dN^{C(P)}/dp_{T}d\eta$ are the differential yields in a central (peripheral)
collision, respectively. Nuclear medium effects are expected to be much stronger in central relative to
peripheral collisions, which makes $R_{CP}$ another measure of these effects. If the yield of the process
scales with the number of binary collisions, $R_{CP}$=1. \\

\section{Results}

The data presented here were collected with BRAHMS detector system~\cite{brahms-nim}. BRAHMS consists
of a set of global detectors for event characterization and two magnetic spectrometers, 
the mid-rapidity spectrometer (MRS) and the forward spectrometer (FS), which identify 
charged hadrons over a broad range of rapidity and transverse momentum. Collision centrality 
is determined from the charged particle multiplicity measured by multiplicity detectors.
Since BRAHMS is a small solid angle device, the average spectrum is obtained by mapping out
the particle phase space by collecting data with many different spectrometer settings. BRAHMS
is the only experiment from RHIC to perform detailed measurements away from midrapidity.


\begin{figure}
\epsfxsize 16.5cm
\epsffile{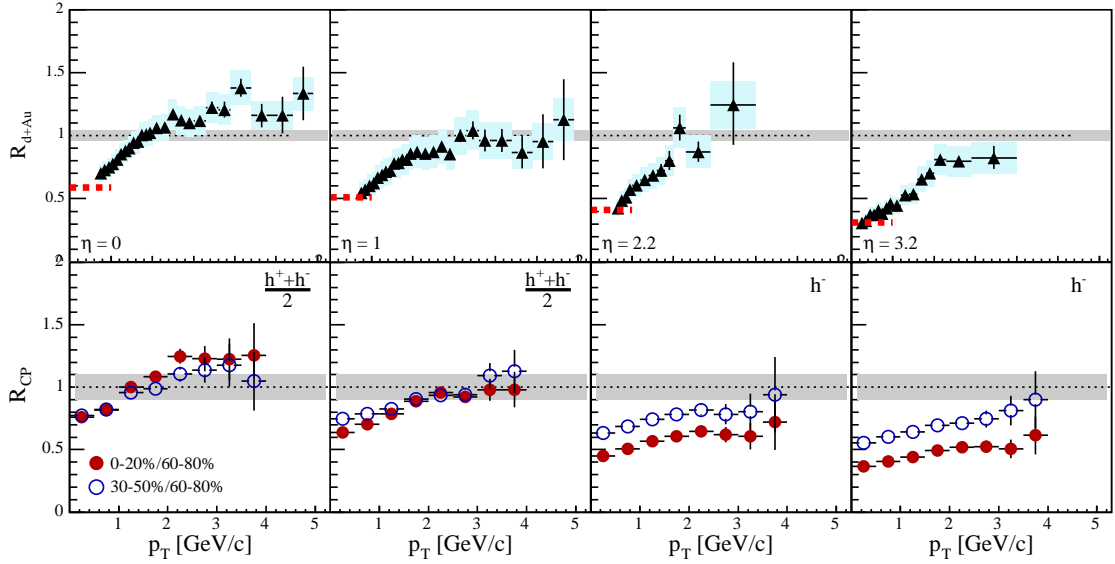}
\caption{Top row: Nuclear modification factor for charged hadrons at $\eta=0, 1.0, 2.2, 3.2$. Systematic 
errors are shown with shaded boxes with widths set by the bin sizes. The shaded band around unity indicates
the estimated error on the normalization to $<N_{bin}>$. Bottom row: Central (red circles) and semi-central
(blue circles) $R_{CP}$ ratios. Shaded bands indicate the uncertainty in the calculation of $<N_{bin}>$ in the
peripheral collisions (12\%). }
\end{figure}

In the absence of high density medium, that is of jet-quenching process, d+Au collisions
are used to study the modifications due to initial state effects. 
Figure 1 shows the pseudorapidity dependence of $R_{dA}$ and $R_{CP}$ for d+Au collisions~\cite{rda}. 
At midrapidity, $R_{dA}$ is showing a Cronin like enhancement with respect to binary scaling limit, for 
transverse momenta greater than 2 GeV/c. This enhancement is thought to be due to the multiple 
scattering of the parton traversing the nucleus prior to the high Q$^2$ scattering that produces the
observed high $p_T$ hadron.
At higher rapidities, the ratio becomes smaller than 1 indicating a suppression in d+Au collisions 
compared to scaled p+p collisions at the same energy, which becomes stronger when going 
to forward angles. 
The bottom row shows the $R_{CP}$ factors as a function of pseudorapidity. At midrapidity
the central(0-20$\%$)-to-peripheral(60-80$\%$) ratio is larger than semicentral(30-50$\%$)-to-peripheral ratio
suggesting also an increased Cronin type multiple scattering effect in the more central collisions. 
Conversely, at forward pseudorapidities the more central ratio is more suppressed, indicating a mechanism for
suppression dependent on the centrality of the collision.

\begin{floatingfigure}[r]{0.5\textwidth}
\begin{center}
\epsfig{file=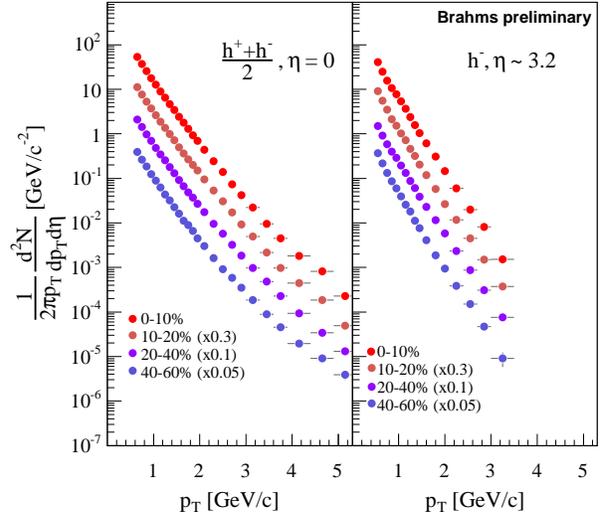,width=9cm,bbllx=2,bburx=550,bbury=420}
\caption{Invariant spectra for charged hadrons produced in Au+Au collisions at $\sqrt{s_{NN}}=200$GeV at 
$\eta=0$ (left panel) and $\eta\sim3.2$ (right panel).}
\end{center}
\end{floatingfigure}

It has been proposed that this suppression at forward rapidity is related to the initial
conditions of the colliding $d$ and Au nuclei, in particular to the possible formation of the Color
Glass Condensate (CGC) in the initial state at RHIC~\cite{cgc}.


For the Au+Au data which will be presented next, the midrapidity spectrometer was positioned
at 90 degrees relative to the beam axis, and measured charged hadrons at pseudorapidities in
the range $\eta < 0.1$. The forward spectrometer was placed at 8 and 4 degrees, for the
ranges in pseudorapidity [2.4, 2.8] and [3.0, 3.5] respectively. The global detectors were
used for the minimum bias trigger and event characterization. This trigger is selecting 
approximately 95\% of the Au+Au interaction cross section. Spectrometer triggers are also
used to enhance the track sample. The IP position is determined with a precision $\sigma <0.85 cm$
by the use of beam counters (BB) placed at $z =\pm 2.2 m$.

Figure 2 shows the measured invariant spectra for inclusive charged hadrons $(h^{+}+h^{-})/2$ at
$90^{0}$ (left panel) and for negative hadrons ($h^{-}$) at $4^{o}$ (right panel), corresponding to
$\eta=0$ and $\eta\sim 3.2$. The displayed spectra are for centralities of 0-10\%, 10-20\%,
20-40\% and 40-60\%. The spectra are from measurements at various magnetic fields (high magnetic 
field chosen in order to increase the statistics at high $p_{T}$) and have been corrected for the
acceptance of the spectrometers and for centrality dependent tracking efficiencies. 
No corrections for feed-down, decay or absorption have been applied for the FS data.

All the charged hadron spectra exhibit the power law shape, where at forward angles
90\% of the particles are emitted in the region $p_{T} <2 GeV/c$.

Figure 3 shows the pseudorapidity dependence of the $R_{CP}$ ratio in Au+Au collisions, 
at $\eta =0$, $\eta\sim 2.6$ and $\eta\sim 3.2$. The observed suppression is
similar at forward rapidities as compared to midrapidity. This result may indicate
that quenching extends in the longitudinal direction.


\section*{Summary}
BRAHMS has measured the rapidity dependence of nuclear modification factors
in d+Au and Au+Au collisions. Away from midrapidity we observe a suppression
of charged hadrons in d+Au collisions, suggesting the enhancement of the
initial state effects in these regions. In Au+Au collisions the suppression
persists over 3 units in pseudorapidity, indicating that the hot and dense
partonic matter could further extend to forward regions.

\section*{Acknowledgments}
This work was supported by the division of Nuclear Physics of the
Office of Science of the U.S. DOE, the Danish Natural Science
Research Council, the Research Council of Norway, the Polish State
Com. for Scientific Research and the Romanian Ministry of
Research.

\begin{figure}
\epsfxsize 16.5cm
\epsffile{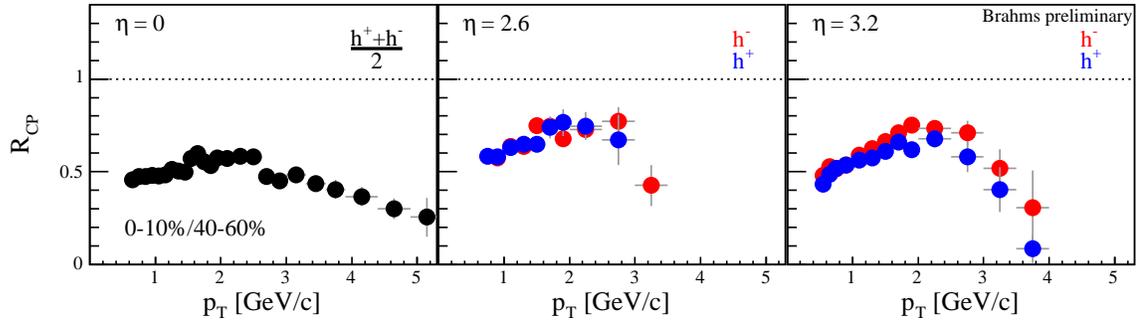}
\caption{$R_{CP}$ ratio for charged hadrons in Au+Au collisions at $\sqrt{s_{NN}}$=200GeV for pseudorapidities 
$\eta=0, 2.6, 3.2$ as a function of transverse momentum $p_{T}$.}
\end{figure}

\end{document}